\documentclass[12pt]{article}

\usepackage[totalheight = 23cm, totalwidth = 17cm]{geometry}
\usepackage{amssymb,amsmath,amsfonts,amsbsy,graphicx}
\usepackage{bm}
\usepackage{color}
\usepackage{subfigure}

\newcommand{\mpl}{m_{\rm Pl}}
\newcommand{\calH}{{\cal H}}
\newcommand{\calL}{{\cal L}}
\newcommand{\calO}{{\cal O}}
\newcommand{\calP}{{\cal P}}
\newcommand{\calR}{{\cal R}}

\begin{document}

\begin{titlepage}

\begin{center}

\rightline{APCTP-Pre2015-025}
\rightline{IPMU15-0196}

\vskip 1cm

\LARGE{\bf Trail of the Higgs in the primordial spectrum}

\vskip 1cm

\large{Jinn-Ouk Gong$^{a,b}$,
Chengcheng Han$^{a,c}$ \, and \,
Shi Pi$^{a,d}$}

\vskip 1.2cm

\small{\it
$^{a}$Asia Pacific Center for Theoretical Physics, Pohang 790-784, Korea
\\
$^{b}$Department of Physics, Postech, Pohang 790-784, Korea
\\
$^{c}$Kavli Institute for the Physics and Mathematics of the Universe,\\
The University of Tokyo, Kashiwa 277-8583, Japan
\\
$^{d}$Institute for Theoretical Physics, Chinese Academy of Sciences, Beijing 100190, China
\\
\vspace{2em}
E-mail: jinn-ouk.gong@apctp.org, chengcheng.han@ipmu.jp, spi@itp.ac.cn
}

\vskip 1.2cm

\end{center}

\begin{abstract}

We study the effects of the Higgs directly coupled to the inflaton on the primordial power spectrum. The quadratic coupling between the Higgs and the inflaton stabilizes the Higgs in the electroweak vacuum during inflation by inducing a large effective mass for the Higgs, which also leads to oscillatory features in the primordial power spectrum due to the oscillating classical background. Meanwhile, the features from quantum fluctuations exhibit simple monotonic $k$-dependence and are subleading compared to the classical contributions. We also comment on the collider searches.

\end{abstract}

\end{titlepage}

\setcounter{page}{0}
\newpage
\setcounter{page}{1}

\section{Introduction}
\label{sec:intro}

The confirmation of the Higgs with a mass $125.09\pm0.21\pm0.11$ GeV at the Large Hadron Collider (LHC)~\cite{Aad:2015zhl} has marked the discovery of the last piece of the jigsaw for the standard model. At the same time, this raises significant questions on the cosmological evolution of the Higgs. If we extend the standard model to higher energy scales, the quartic Higgs coupling becomes negative beyond a scale $\Lambda \sim 10^{10}$ GeV, mainly contributed by the loop corrections from the Yukawa coupling to the top quark~\cite{Sher:1988mj}. At very large field values of the Higgs lies the true minimum with negative Planckian energy density, if extrapolated up to the Planck scale, indicating that the electroweak vacuum is only metastable. This metastability is sensitive to the top quark mass: the current measurement of the top quark mass  $173.34\pm0.76\pm0.3$ GeV~\cite{ATLAS:2014wva} excludes absolute stability at 99\% confidence level~\cite{Degrassi:2012ry}, while the lifetime of the electroweak vacuum is longer than the age of the universe~\cite{EliasMiro:2011aa}. That means, with a generic sub-Planckian initial value of the Higgs $h \lesssim \mpl$, it is very unlikely to find ourselves in the metastable electroweak vacuum $h \lesssim \Lambda \sim 10^{-8}\mpl$ unless the Higgs is forced to stay near the minimum.

Even if the Higgs is initially placed at the origin, during inflation~\cite{inflation} the picture could become different very easily if the energy scale of inflation is large enough. Indeed, with the current bound at 95\% confidence on the tensor-to-scalar ratio $r_{0.05} \lesssim 0.12$~\cite{Ade:2015tva}, it is not difficult to construct inflationary models with the Hubble parameter $H \gtrsim \Lambda$~\cite{Lyth:1998xn}. If the Higgs is only coupled to the standard model particle species, it can be treated as an effectively massless scalar field in the inflationary era. Then the Higgs acquires quantum fluctuations of $\calO(H)$, so that it eventually settles down in the true minimum with large negative energy~\cite{Espinosa:2015qea}.

There are many ways to solve this problem by introducing some new physics beyond standard model. For instance, some new particle with a mass larger than top quark can change the running of the Higgs self-coupling $\lambda$, and prevent it from becoming negative~\cite{EliasMiro:2012ay}. Inflation with finite temperature caused by dissipative effects can also change the effective potential by thermal corrections and raise the vaccum with negative energy to a false one~\cite{Fairbairn:2014zia}. Besides, Higgs inflation with a non-minimal coupling to gravity can solve this problem similarly~\cite{Bezrukov:2014ipa}.

A simple way to make the Higgs safe from instability during inflation is to introduce a large mass by a direct coupling to the inflaton, which makes $m_h \gtrsim H$~\cite{Lebedev:2012sy}. It is shown that the Hawking-Moss decay rate is suppressed by $e^{(m_h/H)^4}$, while the Coleman-de Luccia decay rate is still neglegible~\cite{HM-CdL} so that the evolution of the Higgs towards the electroweak vacuum during inflation is very likely. However, this coupling can affect the evolution of the curvature perturbation during inflation, thus may leave distinctive features primordial power spectrum by resonant oscillations~\cite{resonance,Chen:2011zf}. These potentially important effects have not been studied closely. By treating the coupling term as a perturbation to the standard equation of motion~\cite{Noumi:2013cfa}, we can study in detail the impacts of the heavy Higgs during inflation.

This article is outlined as follows. In Section~\ref{sec:classical}, we first consider the classical background evolution of the Higgs and the inflaton with a simple renormalizable coupling between them. Then we show this oscillating background gives rise to logarithmic oscillations in the primordial spectrum. In Section~\ref{sec:quantum} we compute the contributions from the quantum fluctuations to the power spectrum, which is subdominant compared to the classical features. After briefly commenting on the prospects for collider search in Section~\ref{sec:collider}, we conclude in Section~\ref{sec:conc}.

\section{Background evolution}
\label{sec:classical}

\subsection{Bounds on the Higgs-inflaton coupling}

Let us begin by invoking the Lagrangian with a simple renormalizable term that directly couples the Higgs and the inflaton quadratically,
\begin{equation}
V(\phi,h) = \frac{1}{2}m_\phi^2\phi^2 + \frac{1}{2}\xi\phi^2h^2 + \frac{1}{4}\lambda h^4 \, ,
\end{equation}
where $\phi$ is the inflaton and $\xi$ is a positive coupling constant. For simplicity, we have set the vacuum expectation value of the Higgs to zero, and used the quadratic chaotic potential for inflation. 

We can place an upper bound on $\xi$ by requring that the one-loop quantum correction to the inflaton potential from the Higgs-inflaton coupling
\begin{equation}\label{Vinf1-loop}
\Delta V = \frac{\xi^2\phi^4}{64\pi^2}\log \left( \frac{\xi\phi^2}{m^2} \right)
\end{equation}
is sub-dominant. This gives
\begin{equation}
\label{bound:V<V1-loop}
\xi \lesssim 4\sqrt{2}\pi\frac{m_\phi}{\phi} \approx 9.05\times10^{-6} \left( \frac{A_\calR}{2.19\times10^{-9}} \right)^{1/2} \left( \frac{50}{N} \right)^{3/2} \, ,
\end{equation}
where $A_\calR$ is the amplitude of the power spectrum of the curvature perturbation $\calR$ at the normalization scale and $N$ is the $e$-folding number of that scale from the moment of horizon crossing till the end of inflation.


A lower bound on $\xi$ can be derived from the condition to stabilize the Higgs during inflation. To see this, notice that for a specific value of $\phi$, the maximum along the Higgs direction is obtained from
\begin{equation}
\frac{\partial V}{\partial h} = \xi\phi^2h + \left( \lambda + \frac{\beta_\lambda}{4} \right) h^3 = 0 \, .
\end{equation}
Non-trivial solution only exists for $\lambda+\beta_\lambda/4 < 0$, i.e. $h \gtrsim \Lambda$. For estimation, we may neglect the running of $\lambda$ and take its mimimum value $\lambda_m < 0$ to get
\begin{equation}\label{hmax}
h_\mathrm{max} \approx \sqrt{\frac{\xi}{|\lambda_m|}}\phi \, ,
\end{equation}
which gives the maximum value of the effective potential for a given $\phi$ as
\begin{equation}\label{eq:Vmax}
V_\mathrm{max} = \frac{1}{2}m_\phi^2\phi^2 + \frac{1}{4|\lambda_m|}\xi^2\phi^4 \, .
\end{equation}
Then the probability of the Hawking-Moss decay is $e^{-B_\mathrm{HM}}$ with
\begin{equation}
B_\mathrm{HM} = \frac{2\pi^2}{3|\lambda_m|}\frac{\xi^2\phi^4}{H^4} \, .
\end{equation}
The suppression of this decay rate gives us a lower bound
\begin{equation}
\xi > |\lambda_m|^{1/2}\left(\frac{H}{\phi}\right)^2 \approx 8.64\times10^{-13} \left(\frac{|\lambda_m|}{0.01}\right)^{1/2} \left(\frac{A_\calR}{2.19\times10^{-9}}\right) \left(\frac{50}{N}\right)^2 \, .
\end{equation}
On the other hand, the Coleman-de Luccia decay rate is negligible~\cite{HM-CdL}, similar to the one in the Minkowski spacetime.

\subsection{Pre-inflationary stage}

First we assume that the quartic term of the Higgs is always negligible, otherwise the entire picture is not different with standard model, and the instability catastrophe can not be avoided. Then, we require that the Higgs-inflaton coupling term dominate the potential when $h$ is large (but still sub-Planckian). The universe will not inflate in the beginning for large initial values of $h$, and inflation only happens later when $h$ drops down some critical value and the inflaton potential becomes dominant. From \eqref{eq:Vmax}, if
\begin{equation}\label{bound3}
\xi \gtrsim |\lambda_m|^{1/2}\frac{m_\phi}{\phi} \sim 5.09\times10^{-8} \left(\frac{|\lambda_m|}{0.01}\right)^{1/2} \left(\frac{A_\calR}{2.19\times10^{-9}}\right)^{1/2} \left(\frac{50}{N}\right)^{3/2} \, ,
\end{equation}
the coupling term dominates around $h_\text{max}$ until $h_\text{eq} = m_\phi/\sqrt{\xi}$, at which the inflation potential becomes larger so that inflation begins. If  \eqref{bound3} is violated, then $h_\text{eq} > h_\mathrm{max}$ and the inflaton potential will dominate throughout, leading to standard slow-roll inflation. 

We focus on the former case, and set the initial value of the Higgs close to $h_\mathrm{max}$. The Friedmann equation at this initial condition gives
\begin{equation}
H_\text{max} = \frac{V_\mathrm{max}^{1/2}}{\sqrt{3}\mpl} \approx \frac{\xi\phi_\text{ini}^2}{2\sqrt{3|\lambda_m|}\mpl} \, ,
\end{equation}
where $\phi_\text{ini}$ is the initial value of $\phi$. Then the effective mass of the Higgs, $m_h\equiv\sqrt\xi\phi$,  can be then estimated as
\begin{equation}\label{heavyHiggs}
\frac{m_h}{H} \gtrsim \left\{
\begin{array}{ll}
2\sqrt{\dfrac{3|\lambda_m|}{\xi}} \dfrac{\mpl}{\phi_\text{eq}} \sim 18 & \text{ for } h_\text{eq} < h < h_\text{max} \, ,
\\
\sqrt{6\xi}\dfrac{\mpl}{m_\phi} \sim 76 & \text{ for } h < h_\text{eq} \, .
\end{array}
\right.
\end{equation}
Therefore in both cases the Higgs can be treated as a massive scalar field. The dynamics of the background evolution is then as follows: when released from its initial value near the hilltop of the potential, the Higgs fastly rolls down the potential along the $h$-direction, and oscillates in the valley around the electroweak vacuum. It will be trapped there after around 20 $e$-folds [see \eqref{happrox} and below]. Meanwhile the inflaton is always slowly rolling down the $\phi$-direction as will be discussed soon, and inflation  happens only a few $e$-folds after $h\lesssim h_\text{eq}$. Note that due to rapid oscillations, adopting the simple prescription of integrating out the Higgs without taking into account the derivative operators~\cite{intout} calls for caution.

The pre-inflationary evolution can be treated as a matter dominated universe.
It is convenient to make $\phi$ nearly frozen, which will be justified later soon. The Friedmann equation and the equations of motion for $\phi$ and $h$ are now
\begin{align}
\label{eq:Friedmann}
\frac{1}{2}\dot{h}^2 + \frac{1}{2}m_\phi^2\phi^2 + \frac{1}{2}\xi\phi^2h^2 & = 3\mpl^2H^2 \, ,
\\
\label{eq:phi}
\ddot\phi + 3H\dot\phi + \xi\phi h^2 & = 0 \, ,
\\
\label{eq:h}
\ddot{h} + 3H\dot{h} + \xi\phi^2h & = 0 \, .
\end{align}
Furthermore, we write the quadratic potential for $\phi$ as the Hubble parameter during slow-roll inflation,
\begin{equation}
H_\text{inf} = \frac{m_\phi\phi_\text{eq}}{\sqrt{6}\mpl} \, .
\end{equation}
Then, \eqref{eq:Friedmann} can be rewritten as
\begin{equation}
\dot{h}^2 + (m_hh)^2 = 6\mpl^2 \left( H^2 - H_\text{inf}^2 \right) \, .
\end{equation}
Now we define a new variable $\theta$ as
\begin{align}
  \dot h &= \sqrt{6}\mpl\sqrt{H^2-H_\text{inf}^2}\sin\theta \, ,
  \\
  m_h h &= \sqrt{6}\mpl\sqrt{H^2-H_\text{inf}^2}\cos\theta \, .
\end{align}
Together with (\ref{eq:phi}) we can solve~\cite{Mukhanov:1990me}
\begin{align}
\label{hpre}
h & = h_\text{eq} \text{csch} \left[ \frac{3}{2}h_\text{eq}t \left( 1 - \frac{\sin\left( 2\sqrt\xi\phi_\text{eq} \right)t}{2\sqrt\xi\phi_\text{eq}t} \right) \right]  \cos\left(\sqrt\xi\phi_\text{eq}t\right) \, ,
\\
\label{Hpre}
H & = H_\text{inf} \frac{H_\text{inf} \tanh \left[ \dfrac{3}{2}H_\text{inf} t \left( 1 - \dfrac{\sin\left( 2\sqrt\xi\phi_\text{eq}t \right)}{2\sqrt\xi\phi_\text{eq}t} \right) \right] + H_\text{max}}
{H_\text{inf} + H_\text{max} \tanh\left[ \dfrac{3}{2}H_\text{inf} t \left( 1-\dfrac{\sin\left( 2\sqrt\xi\phi_\text{eq}t \right)}{2\sqrt\xi\phi_\text{eq}t} \right) \right]} \, .
\end{align}
The junction point marks the equality of the two potential terms, which gives $h(t_\text{eq}) \approx h_\text{eq}$ and $H(t_\text{eq}) \approx \sqrt2H_\text{inf}$. These asymptotic solutions,
\begin{align}
h & \approx h_\text{eq} \, \text{csch}\left(\frac{3}{2}H_\text{inf}  t\right) \cos\left(\sqrt\xi\phi_\text{eq}t\right) \, ,
\\
H & \approx H_\text{inf} \coth\left(\frac{3}{2}H_\text{inf} t\right) \, ,
\end{align}
are reliable in the region $\xi^{-1/2}\phi\ll t\ll H_\text{inf}^{-1}$.

Now we turn to see why we can treat $\phi$ as a constant in this stage by calculating the $e$-folding numbers $N_\text{pre}$ elapsed from $h_\text{max}$ until $h_\text{eq}$. Being matter dominated, $N_\text{pre}$ is given by
\begin{equation}\label{Npre}
N_\text{pre} \approx \frac{2}{3}\log \left( \frac{\xi}{2\sqrt{|\lambda_m|}} \frac{\phi_\text{ini}}{m_\phi} \right) \, .
\end{equation}
Using the upper bound \eqref{bound:V<V1-loop}, we can find $N_\text{pre} \lesssim 3.2$. Thus, the range $\phi$ has excursed can be estimated as
\begin{equation}
\frac{\Delta\phi}{\phi_\text{ini}} \sim \left( \frac{3h_\text{eq}}{2\phi_\text{ini}} \right)^2 \ll 1 \, .
\end{equation}
Therefore we can take $\phi$ to be a constant $\phi \approx \phi_\text{ini} \approx \phi_\text{eq}$ during this pre-inflationary stage.

\subsection{Inflationary stage}

After the amplitude of $h$ drops below $h_\text{eq}$, the quadratic inflaton potential begins to dominate and inflation begins soon. Thus we can divide $\phi$- and $h$-parts separately, and solve them perturbatively by treating the cross-term, which becomes subdominant during inflation, as perturbation. First let us solve for pure $\phi$ contributions, the solution for which is the same as in the ordinary inflation dominated by a quadratic potential. Using the slow-roll condition and \eqref{eq:Friedmann}, we find
\begin{align}
\phi & = \phi_\text{eq} - \sqrt{\frac{2}{3}}m_\phi\mpl \left( t-t_\text{eq} \right) \, ,
\\
H_\phi^2 & = H_\text{inf}^2 \left[ 1 + \frac{\epsilon_0}{3} - 2\epsilon_0H_\text{inf} \left( t-t_\text{eq} \right) + \epsilon_0^2H_\text{inf}^2 \left( t-t_\text{eq} \right)^2 \right] \, ,
\end{align}
with the subscript $\phi$ for the Hubble parameter denoting that only $\phi$ contributions are included. Here we have normalized the solution at $t=t_\text{eq}$ and have defined the leading slow-roll parameter $\epsilon_0$ as
\begin{equation}
\epsilon_0 \equiv -\frac{\dot{H}_\phi}{H_\text{inf}^2} = \frac{m_\phi^2}{3H_\text{inf}^2} \, .
\end{equation}

In solving the equation for $h$, we can treat both $H$ and $\phi$ as their zeroth order approximation $H_\text{inf}$ and $\phi_\text{eq}$, and we have
\begin{equation}\label{hinf}
h \propto \exp \left[ -\left( \frac{3}{2} \pm \sqrt{\frac{9}{4} - \frac{\xi\phi_\text{eq}^2}{H_\text{inf}^2}} \right) H_\text{inf}t \right] \sim e^{-3H_\text{inf}t/2}e^{\pm i\mu H_\text{inf} t} \, ,
\end{equation}
where we have defined a mass parameter $\mu \equiv \sqrt{\xi\phi_\text{eq}^2/H_\text{inf}^2 - 9/4} \approx m_h/H_\text{inf}$, as we have $\mu\gg1$ during inflation~[see \eqref{heavyHiggs}]. This solution (\ref{hinf}) should be matched to \eqref{hpre} at $h_\text{eq}$, where
\begin{align}
\label{junction}
h\left(t_\text{eq}\right) & = H_\text{inf} \cos \left( m_ht_\text{eq} \right) \, ,
\\
\label{hinf2}
\dot{h}\left(t_\text{eq}\right) & = -m_\phi\phi_\text{eq} \sin \left( m_h t_\text{eq} \right) - \frac{3}{\sqrt2}H_\text{inf} h_\text{eq} \cos \left( m_ht_\text{eq} \right) \, .
\end{align}
The derivative is continuous only when $\mu\gg1$, and the solution gives a damping oscillation with frequency $m_h$:
\begin{equation}\label{happrox}
h = h_\text{eq} \left( \frac{a_\text{eq}}{a} \right)^{3/2} \cos\left(\mu H_\text{inf} t \right) \, .
\end{equation}
We can see that during inflation the Higgs is damping as $a^{-3/2}$, and after 20 to 25 $e$-folds it will drop from $h_\text{eq}$ to electroweak scale and be trapped there~\cite{Lebedev:2012sy}.

Now we can substitute \eqref{happrox} into \eqref{eq:Friedmann} to compute the Hubble parameter including the cross-term during inflation as
\begin{align}
H^2 & = H_\text{inf}^2 \left\{ 1 + \frac{\epsilon_0}{3} - 2\epsilon_0H_\text{inf} \left( t-t_\text{eq} \right) + \epsilon_0^2H_\text{inf}^2 \left( t-t_\text{eq} \right)^2 \right.
\nonumber\\
& \qquad\qquad \left. + \left(\frac{a_\text{eq}}{a}\right)^3 \left[ 1 + \frac{3}{2\mu}\sin \left(2\mu H_\text{inf} t\right) + \frac{9}{4\mu^2} \cos^2 \left( \mu H_\text{inf} t \right) \right] \right\} \, .
\end{align}
Consequently, the slow-roll parameter $\epsilon$ receives oscillatory corrections as
\begin{equation}\label{epsilonNLO}
\epsilon = \epsilon_0 + 3 \left(\frac{a_\text{eq}}{a}\right)^3 \left[ \cos^2 \left( \mu H_\text{inf} t \right) - \frac{3}{4\mu}\sin \left( 2\mu H_\text{inf} t \right) \right] \, .
\end{equation}
Note that $\epsilon$ becomes smaller than 1 only when $(a_\text{eq}/a)^3\lesssim1/3$, which means inflation only begins around 0.37 $e$-folds after $t_\text{eq}$.
Likewise, we can find the conformal time as
\begin{equation}\label{eq:tauNLO}
\tau = -\frac{1}{aH_\text{inf}} \left[ 1 + \frac{5}{6}\epsilon_0 + \epsilon_0\log \left( \frac{a_\text{eq}}{a} \right) - \frac18\left(\frac{a_\text{eq}}{a}\right)^3 \left\{ 1 - \frac{3\cos\left[2\mu\left( H_\text{inf}t_\text{eq}-\log\left(a_\text{eq}/a\right) \right)\right]}{\mu^2+4} \right\} \right] \, .
\end{equation}
Here we only keep the leading order to $\epsilon_0$ or $\mu^{-1}$.
The inverse relation gives the scale factor as, also to leading order,
\begin{equation}\label{eq:aNLO}
a = -\frac{1}{H_\text{inf}\tau} \left[ 1 + \frac{5}{6}\epsilon_0 + \epsilon_0\log\left( -k_\text{eq}\tau \right) - \frac18\left(-k_\text{eq}\tau\right)^3 \left\{ 1 - \frac{3\cos\left[2\alpha_\text{eq} - 2\mu\log\left(-k_\text{eq}\tau\right)\right]}{\mu^2+4} \right\} \right] \, .
\end{equation}
We have defined a comoving scale $k_\text{eq} \equiv a_\text{eq}H_\text{inf}$ and a constant phase factor $\alpha_\text{eq} \equiv \mu H_\text{inf}t_\text{eq}$. Note that this $k_\text{eq}$ is only a large scale that we use to mark the beginning of inflation. The cosmic and conformal times are related by
\begin{equation}
t = t_\star - \frac{1}{H_\text{inf}}\log\left( -k_\star\tau \right) \, ,
\end{equation}
where $k_\star$ is the comoving wavenumber of the mode that exits the horizon at $t_\star$ and can be conveniently chosen as the normalization scale for the power spectrum amplitude $A_\calR$.

\subsection{Features from oscillating background}

%
Defining
\begin{equation}
 z \equiv \sqrt{2\epsilon}a\mpl \approx
 -\mpl\frac{\sqrt\epsilon_0}{H_\text{inf}\tau} \left\{ 1+\frac{3}{4\epsilon_0}(-k\tau)^3 \left[ 1+\cos(2\alpha_\text{eq}-2\mu\log(-k_\ast\tau)) \right] \right\} \, ,
\end{equation}
the equation of motion for $u\equiv z\mathcal R$ is
\begin{equation}
u_k'' + \left( k^2 - \frac{z''}{z} \right) u_k = 0 \, ,
\end{equation}
with
\begin{equation}\label{z''/z}
\frac{z''}{z} = \frac{2}{\tau^2} \left\{ 1 + \frac{3}{2\epsilon_0}\mu^2\left(-k_\text{eq}\tau\right)^3 \left[ \cos\left(-2\mu\log\left(-k_\star\tau \right) + 2\alpha_\star \right) - \frac{1}{\mu}\sin\left( -2\mu\log\left( -k_\star\tau \right) + 2\alpha_\star \right) \right] \right\} \, .
\end{equation}
The standard solution $u_k^{(0)}$, for which $z''/z = 2/\tau^2$, is
\begin{equation}\label{aux0}
u_k^{(0)} = \frac{1}{\sqrt{2k}} \left( 1-\frac{i}{k\tau} \right)e^{-ik\tau} \, .
\end{equation}
The correction term from the classical oscillation satisfies the linearized equation of motion for $\Delta{u}_k$
\begin{align}
& \Delta{u}_k'' + k^2\Delta{u}_k + \frac{3\mu^2}{\sqrt{2k}\epsilon} k_\text{eq}^3\tau \bigg\{ \cos\left[ -2\mu\log\left( -k_\star\tau \right) + 2\alpha_\star \right]
\nonumber\\
& \left. \hspace{0.3\textwidth} - \frac{1}{\mu} \sin\left[ -2\mu\log\left( -k_\star\tau \right) + 2\alpha_\star \right] \right\} \left( 1-\frac{i}{k\tau} \right) e^{-ik\tau} = 0 \, .
\end{align}
The solution to this equation is
\begin{equation}\label{aux1}
\Delta{u}_k = \frac{3}{8\sqrt{2k}\epsilon_0} \left( \frac{k_\text{eq}}{k} \right)^3 \frac{(\mu-i)^2[i+\mu(1+k^2\tau^2)](-k_\star\tau)^{-2i\mu}}{(1+\mu^2)} \left( 1-\frac{i}{k\tau} \right) e^{-ik\tau + 2i\alpha_\star} + \text{c.c.} \, ,
\end{equation}
where c.c. means the complex conjugate.

Given the leading solution \eqref{aux0} and the correction \eqref{aux1}, we can easily calculate the power spectrum of the curvature perturbation. To linear order in $\Delta{u}_k$,
\begin{equation}\label{def:Pcl}
\calP_\calR = \frac{k^3}{2\pi^2}\lim_{\tau\to0} \left|\frac{u_k(\tau)}{z(\tau)}\right|^2 = \calP^{(0)}_\calR \left\{ 1 + 2k^3 \lim_{\tau\to0} \tau^2\left[ u_k^{(0)}(\tau)\Delta{u}_k^*(\tau) + \text{c.c.} \right] \right\} \, ,
\end{equation}
where the zeroth order power spectrum is the standard result
\begin{equation}\label{eq:P0}
\calP_\calR^{(0)} = \frac{H_\text{inf}^2}{8\pi^2\epsilon_0\mpl^2} \, .
\end{equation}
Substituting (\ref{aux0}) and (\ref{aux1}) into \eqref{def:Pcl}, after some straightforward calculations we find
\begin{equation}\label{result:DeltaPcl}
\frac{\Delta\calP_\calR^\text{(bg)}}{\calP_\calR^{(0)}} = -\frac{3}{2} \frac{\mu}{\epsilon_0} \left( \frac{k_\text{eq}}{k} \right)^3 \left\{ \cos\left[ 2\mu\log \left( \frac{k}{k_\star} \right) + 2\mu N_k + 2\alpha_\star \right] + \frac1\mu\sin\left[ 2\mu\log\left( \frac{k}{k_\star} \right) + 2\mu N_k + 2\alpha_\star \right] \right\} \, .
\end{equation}
Here, the superscript (bg) means it is the correction from the background oscillations and $N_k=\log(-k\tau_e)$ is the $e$-folding number between the moment when the mode of wavelength $1/k$ exits the horizon and the end of inflation. $k_\text{eq}$, as we mentioned before, is a large scale that parameterize the beginning of inflation. 
On the other hand, $k_\star$ is an arbitrary scale we can use to parameterize the cosmic time. Different choices of $k_\star$ just change the phase $\alpha_\star = \sqrt{\xi}\phi_\text{eq}t_\star$. Furthermore, note that \eqref{result:DeltaPcl} has a large prefactor $\mu/\epsilon_0$. To make the perturbation theory reliable, it has to be suppressed by $(k_\text{eq}/k)^3$, which means $k \gg k_\text{eq}$. This is consistent with the fact that inflation happens a few $e$-folds after $t_\text{eq}$, and the wavelength $1/k$ we observe today exits the horizon only after inflation begins.

\section{Features from quantum fluctuations}
\label{sec:quantum}

In this section we calculate the corrections to the power spectrum of the curvature perturbation from the quantum fluctuations of the heavy Higgs during inflation. In the spatially flat gauge, decomposing $\phi(t,\bm{x})=\bar\phi(t)+\delta\phi(t,\bf{x})$ and $h(t,\bm{x})=\bar{h}(t)+\delta{h}(t,\bm{x})$, and expanding the Lagrangian, at quadratic order we have
\begin{equation}
\calL = a^3 \left[ \frac{1}{2}\dot{\delta\phi}^2 - \frac{(\nabla\delta\phi)^2}{2a^2} + \frac{1}{2}\dot{\delta h}^2 - \frac{(\nabla h)^2}{2a^2} - \frac{1}{2}\left( m_\phi^2+\xi\bar h^2 \right)\delta\phi^2 - 2\xi\bar\phi\bar h\delta\phi\delta h - \frac{\xi\bar\phi^2 + 3\lambda\bar h^2}{2}\delta h^2 \right] \, .
\end{equation}
Moving to the interacting picture denoted by a subscript $I$, we can find the free Hamiltonian $\calH_0$ and the interaction part $\calH_I$ coming from the cross-term as
\begin{align}
\label{Hfree}
\calH_0 & = a^3 \left[ \frac{1}{2}\dot{\delta\phi_I}^2 + \frac{(\nabla\delta\phi_I)^2}{2a^2} + \frac{1}{2}m_\phi^2\delta\phi_I^2 + \frac{1}{2}\dot{\delta h_I}^2 + \frac{(\nabla h_I)^2}{2a^2} + \frac12\xi\bar\phi^2\delta h_I^2 \right] \, ,
\\
\label{H2}
\calH_I & = \frac{a^3}{2}\xi\bar h^2\delta\phi_I^2 + \frac{a^3}{2}\lambda\bar h^2\delta h_I^2 + 2a^3\xi\bar\phi\bar h\delta\phi_I\delta h_I \, .
\end{align}
According to the background evolution studied in the previous section, the field first rolls down rapidly along the $h$-direction, and begins to oscillate around its electroweak vacuum. At the same time, it slowly rolls down along the $\phi$-direction, while inflation only happens along this direction a few $e$-folds after $h < h_\text{eq}$. Therefore, the curvature perturbation can be defined in terms of $\delta\phi$ as
\begin{equation}\label{def:Rc}
\calR = \frac{H}{\dot{\bar\phi}}\delta\phi \approx \frac{\delta\phi}{\sqrt{2\epsilon_0}\mpl} \, .
\end{equation}

Next let us quantize the field fluctuations $\delta\phi_I$ and $\delta h_I$. We can proceed in the standard manner by promoting them as operators with separate creation and annihilation operators, satisfying the canonical commutation relations. Then the mode functions of $\delta\phi_I$ and $\delta h_I$, denoted by $u_k$ and $v_k$ respectively, satisfy the linear equations of motion derived from the free Hamiltonian $\calH_0$,
\begin{align}
\label{eom:u}
u_k'' - \frac{2}{\tau}u_k' + k^2 u_k & = 0 \, ,
\\
\label{eom:v}
v_k'' - \frac{2}{\tau}v_k' + \left[ k^2 + \frac{6\xi}{\tau^2}\left(\frac{\mpl}{m_\phi}\right)^2\right]v_k & = 0 \, .
\end{align}
The solutions to \eqref{eom:u} and \eqref{eom:v} are given by linear combinations of the Hankel functions of first and second kind. Requiring that deep inside the horizon $k\gg aH$ they reproduces the Bunch-Davies vacuum solution~\cite{Bunch:1978yq}, we obtain
\begin{align}
\label{mode_u}
u_k & = \frac{H}{\sqrt{2k^3}} ( 1+i k\tau)e^{-i k\tau} \, ,
\\
\label{mode_v}
v_k & = -i \exp \left( -\frac{\pi}{2}\mu + i\frac{\pi}{4} \right) \frac{\sqrt{\pi}}{2} H(-\tau)^{3/2} H^{(1)}_{i\mu} (-k\tau) \, ,
\end{align}
where the mass parameter $\mu \gg 1$ is the same as before.

To calculate the correlation functions including interaction Hamiltonians, it is convenient to use the in-in formalism~\cite{in-in}. The two-point correlation function for $\delta\phi_I$ is, expanded to the second order in the interacting Hamiltonian,
\begin{align}\label{2-correlation}
\left\langle\mathrm{in}\left| \delta\phi_I(\bm{k})\delta\phi_I(\bm{q}) \right|\mathrm{in}\right\rangle & = \left\langle0\left| \delta\phi_I(t,\bm{k})\delta\phi_I(t,\bm{q}) \right|0\right\rangle
\nonumber\\
& \quad + 2\Im \left[ \int^t_{t_0}dt' \left\langle0\left| \delta\phi_I(t,\bm{k})\delta\phi_I(t,\bm{q})H_I(t') \right|0\right\rangle \right]
\nonumber\\
& \quad -2 \Re \left[ \int_{t_0}^tdt_1\int_{t_0}^{t_1}dt_2 \left\langle0\left| \delta\phi_I(t,\bm{k})\delta\phi_I(t,\bm{q})H_I(t_1)H_I(t_2) \right|0\right\rangle \right]
\nonumber\\
& \quad + \int^t_{t_0}dt_1\int^t_{t_0}dt_2 \left\langle0\left| H_I(t_1)\delta\phi_I(t,\bm{k})\delta\phi_I(t,\bm{q})H_I(t_2) \right|0\right\rangle \, ,
\end{align}
where $|\mathrm{in}\rangle$ is the ``in''-vacuum state of the full Hamiltonian, and $|0\rangle$ is the vacuum of the free Hamiltonian \eqref{Hfree}. Using \eqref{def:Rc}, we can connect the two-point correlation function of $\delta\phi_I$ to the power spectrum of the curvature perturbation by
\begin{equation}\label{def:P}
\frac{H^2}{{\dot{\bar\phi}}^2} \left\langle\mathrm{in}\left| \delta\phi_I(\bm{k})\delta\phi_I(\bm{q}) \right|\mathrm{in}\right\rangle = (2\pi)^3\delta^{(3)}(\bm{k}+\bm{q}) \frac{2\pi^2}{k^3}\calP_\calR(k) \, .
\end{equation}
The first term of \eqref{2-correlation} gives the standard power spectrum as
\begin{equation}
\calP_\calR(k) = \frac{H_\text{inf}^2}{8\pi^2\epsilon_0\mpl^2}+\cdots \, ,
\end{equation}
where we have omitted the next-to-leading order corrections to the power spectrum of $\calO(\epsilon)$. This $k$-independent term is $\calP^{(0)}_\calR$ given by \eqref{eq:P0}.

The second term of \eqref{2-correlation} is the contribution from the quadratic $\delta\phi_I$ term in \eqref{H2}. The relevant interaction Hamiltonian is then
\begin{equation}\label{H2phi}
H_I^{(\phi\phi)} = \frac{a^3}{2}\xi\bar h^2 \int \frac{d^3k}{(2\pi)^3} \delta\phi_{\bm{k}}^I\delta\phi_{-\bm{k}}^I \, .
\end{equation}
Meanwhile, the third and fourth terms of \eqref{2-correlation} are mediated by the 
remaining two terms in \eqref{H2},
where the latter case is contributed by 
\begin{equation}\label{H2cross}
H_I^{(\phi h)} = 2a^3\xi\bar\phi\bar h \int \frac{d^3k}{(2\pi)^3} \delta\phi_{\bm{k}}^I\delta h_{-\bm{k}}^I \, .
\end{equation}
This term 
dominates \eqref{H2phi}.

Substituting \eqref{H2phi} and \eqref{H2cross} back into \eqref{2-correlation}, and using the solutions \eqref{mode_u} and \eqref{mode_v}, we can calculate the integrals in \eqref{2-correlation}. It is a straightforward task, and has an analytic result under the assumption $\mu\gg1$. We give the details in Appendix \ref{app:int}, and only ultilize the result to calculate the correction to the power spectrum by the virtue of \eqref{def:P},
\begin{equation}\label{DeltaPQM}
\frac{\Delta\calP_\calR^\text{(q)}}{\calP_\calR^{(0)}} = \frac32\mu\epsilon_0 \left( \frac{k_\star}{k}\right)^3 \left( \frac{\pi^2}{2} + \frac{1}{4\mu^2} \right) \, .
\end{equation}
Note that there is no oscillating part in this leading-order quantum effect, and also it is subdominant compared to the corrections from the background oscillations \eqref{result:DeltaPcl}. We can thus see that the effects from the background oscillations dominate the corrections to the power spectrum.

\begin{figure}[htbp]
\centering
\includegraphics[width=0.4\textwidth, trim = 0 -2em 0 0]{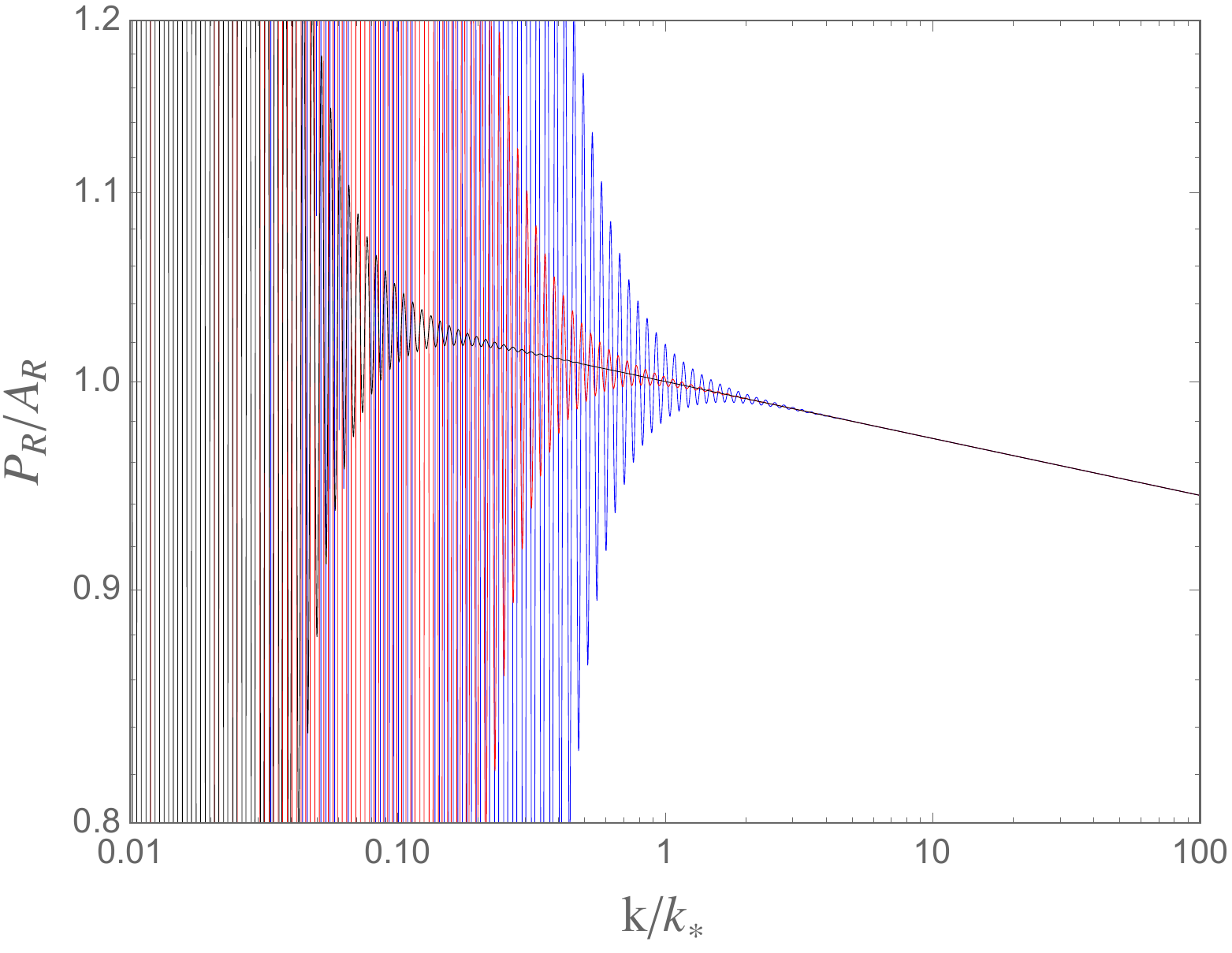}
\includegraphics[width=0.52\textwidth]{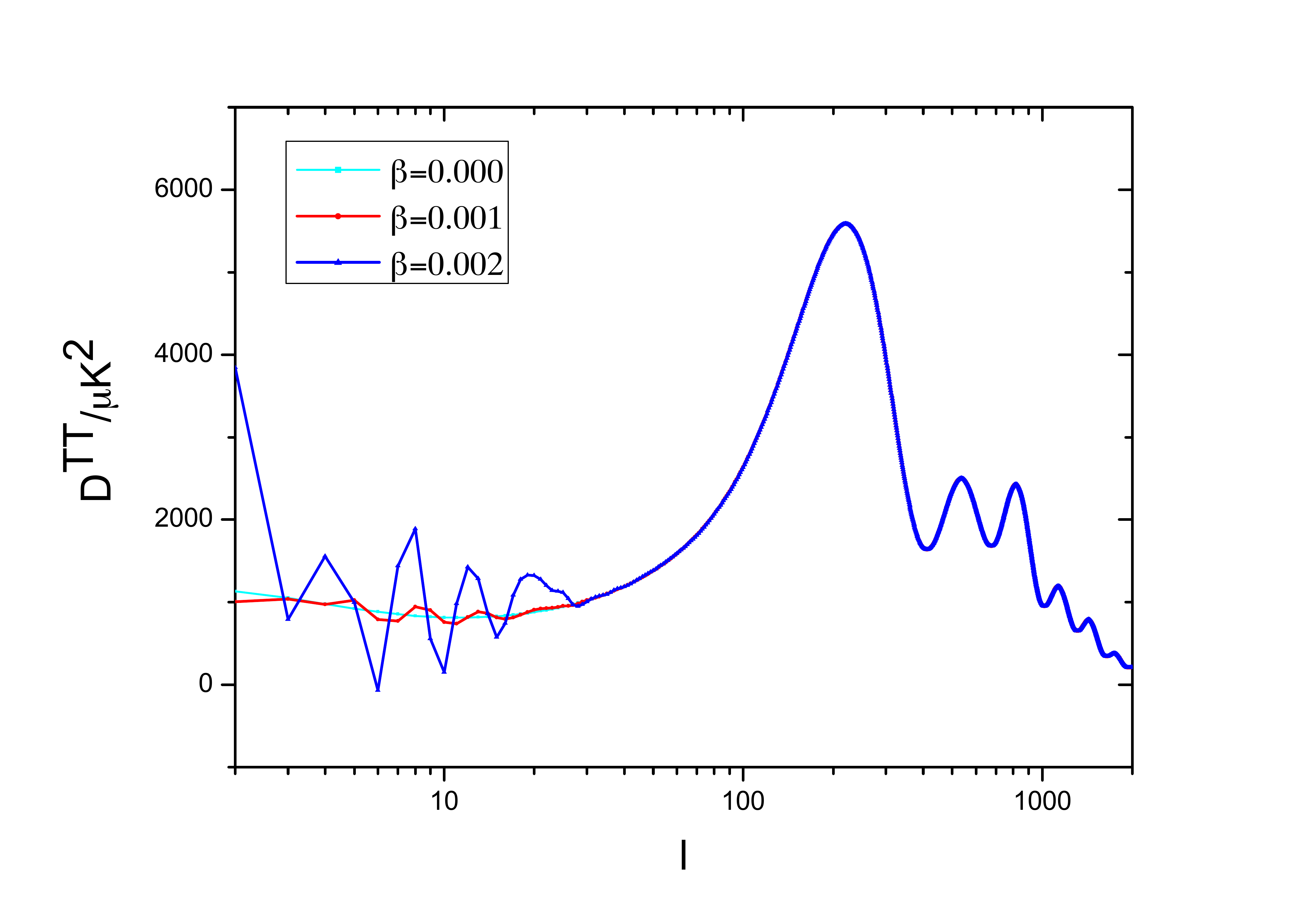}
\caption{
(Left)
the corrected total power spectrum with $\beta = 0.001$ (black), 0.005 (red) and 0.01 (blue). 
(Right) the temperature angular power spectrum of the 
CMB
with $\beta=0$ (cyan), 0.001 (red) and 0.002 (blue).  
}
    \label{fig:spectrum}
\end{figure}

To compare with the Planck data, it is convenient to choose the reference scale $k_\star$ as the pivot scale $0.05\,\text{Mpc}^{-1}$. Since the initial scale $k_\text{eq}$ is arbitrary, let us define a new parameter $\beta \equiv k_\text{eq}/k_\star$. 
To be consistent with the current observation data, we should take $\beta\ll1$, and now we can rewrite the corrections as
\begin{equation}
\frac{\Delta\calP_\calR}{\calP^{(0)}_\calR} = -\frac{3}{2}\frac{\mu}{\epsilon_0} \left(\beta\frac{k_\star}{k}\right)^3 \cos \left\{ 2\mu \left[ \log \left( \frac{k}{k_\star} \right) + N_k + \frac23\sinh^{-1}1 - \log\beta \right] \right\} + \cdots \, ,
\end{equation}
where we have written the leading term only. This results for some choices of $\beta$ depicted in Figure~\ref{fig:spectrum}. We also show the corresponding change in the cosmic microwave background (CMB) spectrum. The cosine of $\log k$ is a typical oscillation pattern originateed from the classical motion like $e^{im_ht}$~\cite{Burgess:2002ub}. The prefactor of $k^{-3}$ reflects both the exponential expansion of the spacetime~\cite{Chen:2011zf} and the quadratic shape of the potential of the heavy field.

\section{Prospects for collider search}
\label{sec:collider}

Once inflation ends, during reheating the energy in the inflaton sector should be transferred into the standard model particles or dark matter. Due to the small coupling with the Higgs, the inflaton itself is hard to be a  candidate of thermal dark matter in our assumptions since during thermal equilibrium the inflaton freezes out and could be over-abundant~\cite{infdm}. If the inflaton is light enough, we may search for possible signatures at colliders. To simplify our analysis, here we only assume that the inflaton mostly decays into dark matter.

After the electroweak symmetry breaking the Higgs acquires the vacuum expectation value $v=246$ GeV, and it induces the interaction $\xi v\phi^2h$ due to the coupling term $\xi\phi^2h^2/2$.
If the inflaton is light enough, $m_\phi<m_h/2$,  the Higgs boson could decay into inflaton pair with the decay width being
\begin{equation}
\Gamma_{h\to\phi\phi} = \frac{\xi^2 v^2}{8 \pi m_h} \sqrt{1-\frac{4 m_\phi^2}{m_h^2}} \, .
\end{equation}
Direct searches which combine the invisible Higgs boson decays in the vector-boson fusion channel and associated production of a Higgs boson with $W$/$Z$ channels set an upper limit on the Higgs boson invisible branching ratio of 23\%~\cite{Aad:2015pla}. Then we can deduce the upper limit of $\xi$ from following formula:
\begin{equation}
\frac{\Gamma_{h\to\phi\phi}}{\Gamma_{h\to\phi\phi} + \Gamma_{h}^\text{SM}} < 23\% \, ,
\end{equation}
which generally gives $\xi<0.008~(0.015)$ for $m_{\phi}= 1~(60)$ GeV. At 3000 $fb^{-1}$ LHC, the limitation of the invisible decay could reach 6\%~\cite{Dawson:2013bba} which corresponds to
$\xi<0.004~(0.007)$ for $m_{\phi}= 1~(60)$ GeV. However at the future $e^+ e^-$ collider, the sensitivity can reach $\sim 0.1\%$, and thus an order of $10^{-4}$ for $\xi$ could be accessed.

For the inflaton mass larger than $m_h/2$, the possible way to probe it is from the inflaton pair production via an off-shell Higgs boson.
At $pp$ colliders, the best channel is vector boson fusion production with a signal of two forward jet plus missing energy. Other possible channels are gluon fusion production with an initial radiate jet, $t\bar{t}$ associated production and $Z$/$W$/$H$-Higgsstrahlung production. Up to now, the limitation from the LHC data
is very weak and there is nearly no limitation for $\xi<1$ with $m_\phi > m_h/2$~\cite{Endo:2014cca}. In the future with a luminosity of 3000 $fb^{-1}$  at 14 TeV LHC (100 TeV SppC)\cite{Craig:2014lda}, the limitation of $m_\phi$ can only reach 130 GeV (150 GeV) for $\xi=1$. So it is very challengeable to probe such a small coupling at future colliders.

\section{Conclusions}
\label{sec:conc}

The Higgs effective potential becomes beyond a energy scale of $\Lambda\sim10^{10}~\text{GeV}$, which raises critical doubt for the stability of the Higgs during inflation. This instability can be cured by introducing a simple direct coupling, such as a quadratic Higgs-inflaton coupling $\xi\phi^2h^2$. The coupling to the  inflaton plays the role of a heavy effective mass of the Higgs during inflation, which can suppress the amplitude of its quantum fluctuation as well as the Hawking-Moss decay rate. We have derived some conditions for which this suppression is effective, which gives a range $8.64\times10^{-13}\lesssim\xi\lesssim9.05\times10^{-6}$.
This coupling is so small that it is very difficult to be observed from collider searches in the near future. However, it can leave distinctive features in the primordial power spectrum of the curvature perturbation during inflation. We have calculated carefully the corrections to the power spectrum due to this Higgs-inflaton cross term, and found that the corrections give rise to typical logarithmic oscillations with a suppression factor sensitive to the initial condition of the beginning of inflation. These corrections can be observed on the largest scales on the CMB, as low-$\ell$ power modulations for instance, if inflation happens only a few $e$-folds before the largest scale we observe today has left the horizon. 
%
%

\subsection*{Acknowledgements}

We would like to thank Qing-Guo Huang and Ying-li Zhang for useful discussions. JG and SP are grateful to the Kavli Institute for Theoretical Physics China for hospitality during the workshop ``Early Universe, Cosmology and Fundamental Physics'', where this work was under progress.
We acknowledge the Max-Planck-Gesellschaft, the Korea Ministry of Education, Science and Technology, Gyeongsangbuk-Do and Pohang City for the support of the Independent Junior Research Group at the Asia Pacific Center for Theoretical Physics.
This work is also supported in part by a Starting Grant through the Basic Science Research Program of the National Research Foundation of Korea (2013R1A1A1006701). CH is supported by World Premier International Research Center Initiative, MEXT, Japan.

\appendix

\section{Some calculations of integrals}
\label{app:int}

In this appendix, we deal with the integrals we have met in \eqref{2-correlation}. One typical integral is $\int x^{a\pm2i\mu}e^{-2ix}$, where $a$ is a constant of $\calO(1)$. It has an analytical result, but what we need is its asymptotic behavior for $\mu\gg1$. To see this, we use the Taylor series of the exponent, take the limit $\mu\gg1$, and then resum the leading order after the integration. We find
\begin{equation}\label{xintegral}
  \int x^{a\pm2i\mu}e^{-2ix}dx=\frac{x^{a+1\pm2i\mu}}{\pm2i\mu}e^{-2ix} \, .
\end{equation}
Usually the range of the integral is taken from 0 to $\infty$. For the upper limit, add an imaginary negative part $i\varepsilon$ to $x$, and the strongly oscillating part in $x\to\infty(1-i\varepsilon)$ converges to 0~\cite{in-in}. The lower limit is also zero for $a>-1$, therefore
\begin{equation}\label{xintegral0}
  \int^\infty_0 x^{a\pm2i\mu}e^{-2ix}dx \to 0 \qquad \text{for} \qquad a>-1 \, .
\end{equation}
If we calculate it analytically, we can see that the result is actually proportional to $e^{-\pi\mu}$~\cite{Chen:2009zp}, which in this article will be treated as 0.

However, there is an infrared divergence for $a\leq-1$. This divergence is spurious because we approximate the inflationary spacetime to be a time-invariant de Sitter one, which is not the case after inflation ends. To include the time-invariance breaking at the end of inflation, the lower limit in \eqref{xintegral} should be replaced by a small infrared cutoff $\tau_e$. This cutoff is connected to the comoving time $\tau_k=-1/k$ when a scale $k$ exits the horizon by
\begin{equation}\label{taue}
  \frac{\tau_e}{\tau_k}=\frac{a_k}{a_e}=e^{-N_k} \, ,
\end{equation}
where $N_k$ is the $e$-folding number from $\tau_k$ to the end of inflation. Therefore, the divergent part is now
\begin{equation}\label{IRdiv}
  \lim_{x\to0}\log x \to \log(-k\tau_e)=\log \left( \frac{\tau_e}{\tau_k} \right) + \log(-k\tau_k)=-N_k \, ,
  \end{equation}
where for the last step we have used $k\tau_k=-1$. Then from \eqref{xintegral} we have
\begin{equation}\label{xintegralIR}
\int^\infty_0dx~x^{a\pm2i\mu}e^{-2ix} \to e^{-N_k(a+1)}\frac{e^{\mp2i\mu N_k}}{\mp2i\mu} \qquad \text{for} \qquad a\leq-1 \, .
\end{equation}
We see that this integral is exponentially enhanced, unless $a=-1$ for which we have
\begin{equation}\label{ExpN}
  \int^\infty_0dx~x^{-1\pm2i\mu}e^{-2ix} \to \frac{e^{\mp2i\mu N_k}}{\mp2i\mu} \, .
\end{equation}
Now the large number $N_k$ only appears in the phase, to which the result is not sensitive.

Another useful integral is~\cite{Gong:2001he}
\begin{equation}\label{Ei}
  \int^\infty_xdy~\frac{e^{-iby}}{y}=-i\pi-\text{Ei}(-ibx) \, .
\end{equation}
We put the upper limit $y\rightarrow\infty(1-i\varepsilon)$ to average the highly oscillating term.
If we make the lower limit to be 0, and use \eqref{IRdiv} to remove the spurious infrared divergence, we have
\begin{equation}\label{Ei2}
  \int^\infty_0dy~\frac{e^{-iby}}{y}=\lim_{x\rightarrow0}\log(bx)+\gamma-\frac{i\pi}{2}=N_k-\gamma-\log b-\frac{i\pi}{2} \, ,
\end{equation}
where $\gamma$ denotes the Euler-Mascheroni constant.

To calculate the third term in \eqref{2-correlation}, we have to use the asymptotic form for the Hankel function of the first kind when the order is pure imaginary and large,
\begin{equation}\label{Hankelasym}
  H_{i\mu}^{(1)}(x)\rightarrow\sqrt{\frac{2}{\pi\mu}}e^{\pi\mu/2}x^{i\mu}\times\mathrm{phase} \, ,
\end{equation}
where ``phase'' means some constant phase term $e^{i\phi}$ which will be eliminated by the $H_{i\mu}^{(1)\ast}$ term in the inner integral. Therefore we have
\begin{align}
\label{B3}
& \Re \left[ \int_{t_0}^tdt_1\int_{t_0}^{t_1}dt_2 \left\langle0\left| \delta\phi_I(t,\bm{k})\delta\phi_I(t,\bm{q})H_I(t_1)H_I(t_2) \right|0\right\rangle \right]
\nonumber\\
= & (-i)^2\mu\epsilon_0\frac{3H_\text{inf}^2}{2k^3}\left(\frac{k_\star}{k}\right)^3 \Re \left[ e^{-2i\alpha_\star} \left(\frac{k_\star}{k}\right)^{2i\mu} \beta_1 + \beta_2 + \beta_3 + e^{2i\alpha_\star} \left(\frac{k_\star}{k}\right)^{-2i\mu}\beta_4 \right] \, ,
\\
\beta_1 \equiv & \int^\infty_0 dx \left(\frac1x+i\right)x^{2i\mu} e^{-ix} \int^\infty_x dy \left(\frac1y+i\right)e^{-iy} \, ,
\\
\beta_2 \equiv & \int^\infty_0 dx \left(\frac1x+i\right)x^{2i\mu} e^{-ix} \int^\infty_x dy \left(\frac1y+i\right)y^{-2i\mu}e^{-iy} \, ,
\\
\beta_3 \equiv & \int^\infty_0 dx \left(\frac1x+i\right) e^{-ix} \int^\infty_x dy \left(\frac1y+i\right) e^{-iy} \, ,
\\
\beta_4 \equiv & \int^\infty_0 dx \left(\frac1x+i\right) e^{-ix} \int^\infty_x dy \left(\frac1y+i\right) y^{-2i\mu} e^{-iy} \, .
\end{align}
%
%
Use the integrals \eqref{ExpN} and \eqref{Ei2}, we can easily find
\begin{align}
\label{beta2}
\beta_2 & = \frac{1}{2\mu} \left[-\frac{\pi}{2} + i \left(-N_k + \gamma + \log2 - \frac34 \right) \right] \, ,
\\
\label{beta4}
\beta_4 & = \frac{-e^{2i\mu N_k}}{4\mu^2} \, .
\end{align}
The integrand in $\beta_3$ is symmetric to $x$ and $y$, and we can simplify it by expand the integral area to the entire first quadrant on the $x$-$y$ plane as
\begin{equation}
\label{beta3}
\beta_3 = \frac12 \left[ \int^\infty_0 dx \left(\frac1x+i\right) e^{-ix} \right]^2 = \frac12 \left( N_k+1-\gamma \right)^2 - \frac{\pi^2}{8} - i\frac\pi2(N_k+1-\gamma) \, .
\end{equation}
The integral $\beta_1$ needs a little attention. Calculating the integral for $y$, we find
\begin{equation}
\beta_1 = \int^\infty_0 dx \left(\frac1x+i\right) x^{2i\mu} e^{-ix} \left[ -i\pi-\text{Ei}(-ix)+e^{-ix} \right] \, .
\end{equation}
Using the series expansion 
\begin{equation}
\text{Ei}(-ix)=\gamma-i\frac\pi2+\log x+\sum_{n=1}^\infty\frac{(-ix)^n}{n!\cdot n} \qquad \text{for} \qquad x\neq0 \, ,
\end{equation}
$\beta_1$ can be written as
\begin{align}
\beta_1 & = \int^\infty_0dx\left(\frac1x+i\right)x^{2i\mu}e^{-ix} \left[ -i\frac\pi2-\gamma-\log x-\sum_{n=1}^\infty\frac{(-ix)^n}{n!\cdot n}+e^{-ix}\right]
\nonumber\\
& \equiv \beta_1^{(1)}+\beta_1^{(2)}+\beta_1^{(3)}+\beta_1^{(4)} \, ,
\end{align}
where
\begin{align}
\label{beta11}
\beta_1 & \equiv \left( -i\frac\pi2-\gamma \right) \int^\infty_0 dx \left(\frac1x+i\right) x^{2i\mu} e^{-ix} = \left(-i\frac\pi2-\gamma\right) \frac{e^{-2i\mu N_k}}{-2i\mu} \, ,
\\
\label{beta14}
\beta_4 & \equiv \int^\infty_0 dx \left(\frac1x+i\right) x^{2i\mu} e^{-2ix} = \frac{e^{-2i\mu N_k}}{-2i\mu} \, .
\end{align}
The integral involving the logarithm can be done analytically,
\begin{align}
\beta_1^{(2)} & \equiv -\int^\infty_0 dx \left(\frac1x+i\right) x^{2i\mu} \log x e^{-ix}
\nonumber\\
& = \left. -x^{2i\mu} \left[ \frac{ix}{(2\mu-i)^2}+\frac{1}{4\mu^2}+\left(-\frac{i}{2\mu}+\frac{x}{2\mu-i}\right)\log x \right] \right|^\infty_0 \, .
\end{align}
As usual, modifying the upper limit as $x \to \infty(1+i\varepsilon)$ will ensure it is zero. And the lower limit is $x\rightarrow e^{-N_k}$. Neglecting all the terms that are suppressed by $e^{-N_k}$, we have
\begin{equation}\label{beta12}
\beta_1^{(2)} = \left(\frac{1}{4\mu^2}+\frac{iN_k}{2\mu}\right)e^{-2i\mu N_k} \, .
\end{equation}
The integral involving the infinite sum can be written as a summation of integrals,
\begin{equation}
\beta_1^{(3)} = \sum_{n=1}^\infty \frac{(-i)^n}{n\cdot n!} \int^\infty_0 dx \left(\frac1x+i\right) x^{n+2i\mu}e^{-ix} \, .
\end{equation}
Considering \eqref{xintegralIR}, and also noticing that the summation begins from $n=1$, we immediately see that
\begin{equation}\label{beta13}
  \beta_1^{(3)}=0 \, .
\end{equation}
Now collecting the results of \eqref{beta11}, \eqref{beta12}, \eqref{beta13}, \eqref{beta14}, \eqref{beta2}, \eqref{beta3} and \eqref{beta4}, and substituting them into \eqref{B3}, we finally have
\begin{align}\label{B4}
& \Re \left[ \int_{t_0}^tdt_1\int_{t_0}^{t_1}dt_2 \left\langle0\left| \delta\phi_I(t,\bm{k})\delta\phi_I(t,\bm{q})H_I(t_1)H_I(t_2) \right|0\right\rangle \right]
\nonumber\\
= & -\mu\epsilon_0\frac{3H_\text{inf}^2}{4k^3} \left(\frac{k_\star}{k}\right)^3 \left[ \frac{\pi}{2\mu}\cos\Theta + \frac{N_k-\gamma+1}{\mu}\sin\Theta - \frac{\pi}{4\mu} + (N_k+1-\gamma)^2 - \frac{\pi^2}{4} \right] \, ,
\\
\Theta \equiv & 2\mu \left[ \log\left(\frac{k}{k_\star}\right)+N_k+\alpha_\star \right] \, .
\end{align}

The calculations of the second and fourth terms in \eqref{2-correlation} are similar, and we omit the process and give only the results:
\begin{align}
\label{A4}
& \left[ \int^t_{t_0}dt' \left\langle0\left| \delta\phi_I(t,\bm{k})\delta\phi_I(t,\bm{q})H_I(t') \right|0\right\rangle \right] = \epsilon_0 \frac{3H_\text{inf}^2}{4k^3} \left(\frac{k_\star}{k}\right)^3 \left(-\frac\pi2\right) \, ,
\\
\label{C4}
& \int^t_{t_0}dt_1\int^t_{t_0}dt_2 \left\langle0\left| H_I(t_1)\delta\phi_I(t,\bm{k})\delta\phi_I(t,\bm{q})H_I(t_2) \right|0\right\rangle
\nonumber\\
= & \mu^2\epsilon_0 \frac{3H_\text{inf}^2}{4k^3} \left(\frac{k_\star}{k}\right)^3 \left[ (N_k+\gamma-1)^2 + \frac{\pi^2}{4} + \frac{N_k+1-\gamma}{\mu}\sin\Theta + \frac{\pi}{2\mu}\cos\Theta + \frac{1}{4\mu^2} \right] \, ,
\end{align}
%
%
which, together with \eqref{B4}, give the corrections to the power spectrum via the quantum effects \eqref{DeltaPQM}.
%
%
An interesting fact is that there is no oscillating part in this quantum effect, and also it is subleading compared to the effect from the background oscillation \eqref{result:DeltaPcl}.

\end{document}